\documentclass[10pt,twocolumn]{article}
\usepackage{amsmath,amssymb}
\usepackage{citesort,epsfig}
\begin{document}

\title{Flory-Huggins theory for athermal mixtures of hard spheres and larger
flexible polymers}
\author{{\bf Richard P. Sear}\\
Department of Physics, University of Surrey,
Guildford, Surrey GU2 7XH, United Kingdom\\
{\tt r.sear@surrey.ac.uk}}
\maketitle

\begin{abstract}
A simple analytic theory for mixtures of hard spheres and larger
polymers with excluded volume interactions is developed.
The mixture
is shown to exhibit extensive immiscibility. For large polymers
with strong excluded volume interactions, the
density of monomers at the critical point for demixing
decreases as one over the square root of the
length of the polymer, while the density of spheres tends to
a constant. This is very different to the behaviour of
mixtures of hard spheres and ideal polymers, these mixtures although even
less miscible than those with polymers with excluded volume interactions,
have a much higher polymer density at the critical point of demixing.
The theory applies to the complete range of mixtures
of spheres with flexible polymers,
from those with strong excluded
volume interactions to ideal polymers.
\end{abstract}

\section{Introduction}

Athermal mixtures are remarkable because the only energy they possess
is thermal energy; their behaviour including their phase behaviour
is determined solely by entropy. But this does not mean that this
phase behaviour is uninteresting, entropy alone can drive phase transitions
such as freezing and, in mixtures, demixing into coexisting fluid phases.
This demixing is found almost whenever the two components of the
mixture are very different, such as the athermal polymer and hard spheres
considered here. We look at mixtures of hard spheres and athermal
polymer where the polymer is as large or larger than the spheres:
the mean-square end-to-end separation of the polymer greater than
or equal to the diameter of the spheres. This complements earlier
work on mixtures where the polymer is smaller. As might be expected we
find extensive immiscibility between the polymer and the spheres.
We calculate phase boundaries and determine their scaling with the sizes
of the components, for example, the density of monomers at the critical
point is found to scale as one over the square root of the number
of monomers in a polymer, just as does the critical point for demixing
of a polymer and a poor solvent.

We are most interested in the limit where the size of the polymer,
$R_E$, is much greater than that of the colloid, $\sigma$.
$R_E$ is the root-mean-square end-to-end distance of the polymer,
and $\sigma$ is the hard-sphere diameter.
Our theory is motivated by the idea that for $R_E\gg\sigma$ the effect
of the colloids is a small length-scale effect in the sense that if the
colloid degrees of freedom of the mixture are integrated out
we are left with polymers which at length-scales large
in comparison to $\sigma$ behave qualitatively just like a polymer in
structureless solvent. The colloids effect the quality of this
effective structureless solvent but do not introduce any new physics
on length-scales large with respect to $\sigma$.
If this is correct
then the long length-scale behaviour and the phase behaviour, of
colloid-polymer mixtures may be mapped onto the well-understood
long length-scale and
phase behaviour of a polymer plus solvent system. The phase separation of
polymer and much smaller colloidal spheres will be qualitatively
identical to that of a long polymer and a poor solvent. This
phase separation has been extensively studied and is quite well
understood \cite{degennes,frauenkron97,wilding96,yan00}.
The approximate theory we will develop is essentially
a Flory-Huggins polymer-plus-solvent free energy with an
effective solvent quality which depends on the concentration of
the colloid. It is known that a Flory-Huggins free energy describes
rather well the qualitative features of
the phase separation of a long polymer and a poor solvent;
see the comparison with computer simulation data in
Refs.~\cite{frauenkron97,wilding96,yan00}. It is also known that
purely entropic effects can result in the polymer effectively being
in a poor solvent \cite{frenkel92}.

The literature on mixtures of colloidal particles and non-adsorbing polymers
is extensive because colloid-polymer mixtures are common, and the
limit we consider where all interactions are purely repulsive is
a rather fundamental limit of these mixtures.
Mixtures
in which the polymer molecules are both larger than the particles and
flexible (as opposed to semiflexible) are
formed when the particles are small, a few nms across. Nanoparticles
are colloidal particles of this size as are proteins. Protein-polymer
mixtures are common, for example polymers are mixed with proteins
in order to induce the proteins to crystallise \cite{durbin96,piazza00}.
Although in practice it
is unlikely that the monomer-protein interaction is ever purely
repulsive over all the surface of a protein, however the limit
we study here provides a basis for incorporating the effects of
weak adsorption of the polymer onto a protein molecule.
The limit in which all three interactions, sphere-sphere, monomer-monomer
and sphere-monomer, are purely repulsive is an important and fundamental
limit. It is fundamental in the sense that as we are assuming the
monomers to be much smaller than the spheres, then the details of the
monomer-sphere interaction are irrelevant as long as it is purely
repulsive; the details of the monomer-monomer interaction only effect
the behaviour by altering a single parameter, the monomer-monomer
second virial coefficient, and any sharply repulsive sphere-sphere
interaction will behave almost like hard spheres. Thus many of the
details of the mixture are irrelevant.

The interactions between colloidal spheres and polymer molecules have been
studied theoretically via a number of techniques: scaling approaches
\cite{degennes79,sear98,sear01a},
field theory \cite{eisenriegler96,hanke99,eisenriegler00,maassen01},
computer simulation \cite{meijer94,louisxxx},
integral equations \cite{fuchs00,fuchs02}, and other approaches
\cite{jansons90,odijk97,odijk00,tuinier00}.
The phase behaviour has been studied via computer simulation
\cite{meijer94}, scaling theory \cite{sear97}, and
perturbation theory \cite{sear01,ramakrishnan02},
where Refs.~\cite{meijer94,sear01}
are for ideal polymers.
Reference \cite{fuchs02} is a review of recent work on colloid-polymer
mixtures, focusing mainly on the structure and on the results of
integral equations, but also discussing other approaches
and the phase behaviour.
Earlier work by the author \cite{sear97}
assumed that the phase separation would occur when the polymer
was semidilute. The more careful and better founded work here finds
that this is not correct, the phase separation occurs at the boundary
between the dilute and semidilute regimes where the free energy
expression assumed in Ref.~\cite{sear97} is not valid. Therefore,
in particular the findings of Ref.~\cite{sear97} for the polymer density where 
phase separation occurs are incorrect and should be discounted.
The opposite limit to that of interest here, i.e., where the
polymer molecules are smaller than the colloidal spheres, has
been considered extensively, see
Refs.~\cite{joanny79,gast83,lekkerkerker92,meijer94,dijkstra99}
and references therein. In this limit the polymer (with or without
monomer-monomer excluded volume interactions) induces crystallisation of the
spheres, there is no equilibrium separation into coexisting fluids.

\begin{figure}[t]
\begin{center}
\caption{
\lineskip 2pt
\lineskiplimit 2pt
A schematic of our mixture of large polymer molecules and
colloidal spheres. The black discs
represent the colloids and the curve represents a polymer coil.
The rescaled monomers used to estimate the polymer-colloid interaction
are drawn as dashed circles.
\label{model}
}
\vspace*{0.1in}
\epsfig{file=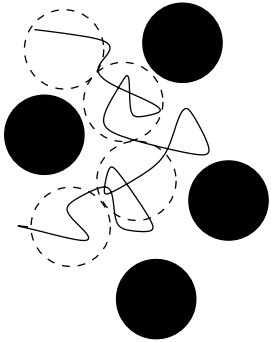,width=1.4in}
\end{center}
\end{figure}

\section{Model and physical picture}

Our colloidal particles are modeled
by hard spheres of diameter $\sigma$
and our polymers are modeled by flexible chains of $N$ monomers,
each of length $a$. We characterise the interaction between
a pair of monomers of the polymer with a second virial coefficient
for this interaction, $B_M$. We start with a polymer in a good
solvent far from the theta temperature where $B_M=O(a^3)$ and
the polymer behaves as a self-avoiding walk (SAW).
The interaction between a pair
of hard spheres is similarly characterised by its second virial
coefficient $B_S=(2/3)\pi\sigma^3$. 
The interaction between a sphere and a monomer is taken to be purely
repulsive and short range, a monomer cannot overlap with a sphere
but otherwise does not interact with it. The polymers do not
adsorb onto the surface of the spheres. 
The monomer size $a$ is much smaller than the
diameter of the spheres, and
here we are considering particles which are smaller than the polymers so
our three length-scales satisfy $a\ll\sigma<R_E$.

A mixture of spheres and polymer is difficult to deal with
because of the sphere-polymer interaction. Pure hard spheres
and pure polymers have both been studied extensively and the
equilibrium behaviour of both is well understood. The
sphere-polymer interaction is hard to deal with because of
the large difference between the size of a sphere and that of a monomer
and because of the
connectivity of the polymer --- if the interaction of
a sphere and a monomer was independent of the interaction of the
sphere with the next monomer along the polymer,
then the sphere-polymer interaction
could be estimated easily. However, when the monomers are much
smaller than the sphere this is very far from being true.
The problem of the disparity in sizes
can be solved by rescaling the monomer size from $a$ to $\sigma$, i.e.,
by viewing the polymer not as being composed of $N$ monomers of
size $a$ but of $n_B$ monomers of size $\sigma$. This rescaling is
quite common, see the book of de Gennes \cite{degennes}. It has
already been applied to mixtures of ideal large polymers and spheres
\cite{sear01}, where it was referred to as the extended
Asakura-Oosawa model. We call the
rescaled monomers of size $\sigma$, blobs.
In principle
this rescaling can be done exactly, i.e., it can be done leaving the
free energy etc.~exact, but here we view it as part
of a physically motivated approximation scheme.
See Fig.~\ref{model} for a schematic of our sphere-polymer mixture
showing the polymer composed of a chain of blobs of size $\sigma$.
If there are $N_B$ monomers of size $a$ in one blob of size $\sigma$
then $n_B$ is related to $N$ by $n_B=N/N_B$.
So, relating the number of blobs to the number of monomers
requires estimating $N_B$, we defer this
to section \ref{secsq}. Until then we specify the polymer size
by specifying $n_B$ and do not concern ourselves with how $n_B$ is
found for a given polymer.
The approximate free energy we will obtain
depends on the polymer length only through $n_B$.

Having performed the rescaling we approximate the interaction between a sphere
and a polymer as being $n_B$ independent sphere-blob interactions.
This is quite a reasonable approximation as the blobs and spheres
are of the same size. Applying this approximation before rescaling,
i.e., approximating the sphere-polymer interaction by $N$ sphere-monomer
interaction is qualitatively wrong because the monomers are so
much smaller than the sphere so the sphere interacts with many monomers
at a time. We rescaled the monomers just so we could apply this
simplifying approximation. The blob-blob and
sphere-blob interactions are
characterised by their second virial coefficients,
$B_B$ and $B_{SB}$ respectively.
%The second virial coefficient
%for the interaction between a sphere and a blob is approximated by
%that for the interaction between two hard spheres of diameter $\sigma$
%\begin{equation}
%B_{SB}=\frac{4}{3}\pi\sigma^3=2B_S,
%\label{bsp}
%\end{equation}
%i.e., a blob of size $\sigma$ in effect excludes
%a sphere of diameter $\sigma$
%from a spherical volume of radius $\sigma$.
The second virial coefficient
for the interaction between a sphere and a blob is of order that
for the interaction between two hard spheres of diameter $\sigma$
but is a little smaller. We defer its estimation to section \ref{secsq}.
Returning to our assumption that
the interaction between a polymer and a sphere consists of $n_B$
independent blob-sphere interactions, this implies that the
second virial coefficient for the sphere-polymer interactions 
is $n_BB_{SB}$. Finally, we remark that after rescaling, our mixture
of polymers of blobs and spheres resembles the athermal polymer + solvent
mixture considered by Frenkel and Louis \cite{frenkel92}. The phase
separation in both cases is driven by unfavourable excluded volume
interactions.

\section{Flory-Huggins--type theory}

We start with the basic Flory-Huggins theory for a polymer of monomers
which interact only via excluded volume interactions,
see for example the book of de Gennes, pages 113-115.
This is often referred to as an athermal polymer solution.
For our `monomers' we take the blobs not the original monomers of length $a$.
Thus the `monomer' density in the theory is actually the density of
blobs which is equal to $\rho_Pn_B$;
$\rho_P$ is the number density of polymer molecules.
As usual we use a reduced
`monomer' density, $\phi$,
which we obtain by multiplying the blob number density
by the volume one blob excludes to another, $2B_B$, $\phi=\rho_Pn_B(2B_B)$.
The Flory-Huggins Helmholtz free energy $F$ then has the usual form
\begin{equation}
\frac{F(2B_B)}{V}=f= \frac{\phi}{n_B}\ln\phi+(1-\phi)\ln(1-\phi),
\label{flory}
\end{equation}
which defines the reduced Helmholtz free energy per unit volume $f$.
Throughout, we use units such that the thermal energy $kT=1$.
This is for a polymer solution, no colloidal spheres present. We add
on the contribution of the colloidal spheres using a virial expansion,
\begin{eqnarray}
f&=&\frac{\phi}{n_B}\ln\phi+(1-\phi)\ln(1-\phi)+\nonumber\\
&&(2B_B)\left\{ \rho_C\left[\ln\rho_C-1\right]+
\rho_C^2B_S+\rho_C\rho_Pn_BB_{SB}
\right\},\nonumber\\
\label{flory_mix}
\end{eqnarray}
where we have truncated the expansion after the second virial
coefficient terms. We have dropped cubic and higher order terms,
which is only valid at low colloid densities.
The last term within the braces is the second virial coefficient
term for the polymer-sphere interaction: within our approximation
it is just $n_B$ independent sphere-blob excluded volume interactions,
each with an excluded volume $B_{SB}$.
In order to obtain a simple analytic theory
we will neglect not only all terms for the sphere-polymer interaction
beyond the leading order,
second virial coefficient term,
but all the terms from sphere-sphere interactions, including the
leading order $\rho_C^2B_S$ term shown in Eq.~(\ref{flory_mix}). This
latter approximation is quite severe but we do this in the expectation
that the sphere-sphere interactions will not be very large when the
colloidal spheres and polymer demix and that the sphere-sphere
interactions, unlike the blob-blob interactions, are not essential
to understanding the basic physics of this demixing.

Making these two approximations and defining a reduced density of
spheres $\phi_C=\rho_C(2B_B)$, Eq. (\ref{flory_mix}) becomes
\begin{equation}
f=\frac{\phi}{n_B}\ln\phi+(1-\phi)\ln(1-\phi)+
\phi_C\left[\ln\phi_C-1\right]+\phi_C\phi b,
\label{flory_mix2}
\end{equation}
where we have changed the $\ln\rho_C$ term to $\ln\phi_C$ term which
we can do as the difference between the two is a constant, $\ln(2B_B)$,
which has no effect on the phase behaviour. The quantity $b$ is the ratio
between the excluded volumes of the sphere-blob and blob-blob
interactions, $b=B_{SB}/(2B_B)$.

Apart from a somewhat more
complex dependence on $\phi$, Eq.~(\ref{flory_mix}) is of the same
form as the free energy of Eq.~(1) of Ref.~\cite{sear95}.
Although in that work the mixture
was of a mixture of thick and thin hard rods and the free energy was
exact.  Below,
we will transform Eq.~(\ref{flory_mix2}) following the same approach
as used in Ref.~\cite{sear95}.
The free energy Eq.~(\ref{flory_mix2}) is linear in the density of
spheres, $\phi_C$. This makes it easy to analytically transform from
fixed $\rho_P$ and $\rho_C$ to fixed $\rho_P$ and $\mu_C$, where
$\mu_C$ is the chemical potential of the colloidal spheres. This
transform is useful as then we have a thermodynamic potential, called
the semigrand potential, which depends on one density variable,
$\rho_P$ or $\phi$, and one field variable, $\mu_C$. This is
completely analogous to the Helmholtz free energy of a single component
system in which temperature is important; that free energy
also depends on a density variable
(the number density) and a field variable (the temperature).
As such once we have the semigrand
potential calculating phase equilibria is just as easy as for a single
component system.

So, at fixed $\phi$ and $\mu_C$ the relevant thermodynamic function
is the semigrand potential $\Omega$. In
fact it is slightly more convenient to work
with the activity of the colloid $z_C=\exp(\mu_C)$ not the
chemical potential.
We need the semigrand potential $\Omega$ which is a Legendre
transform of the Helmholtz free energy
\begin{equation}
\frac{\Omega(2B_B)}{V}=\omega=f-\phi_C\mu_C,
\label{omeg}
\end{equation}
which defines the reduced semigrand potential per unit volume
$\omega$ \cite{notesg}.
The chemical potential $\mu_C$ is just the $\phi_C$ derivative of $f$, so,
taking this derivative of Eq.~(\ref{flory_mix2}),
\begin{equation}
\mu_C=\ln\phi_C+\phi b=\ln z_C,
\label{muc}
\end{equation}
which can be rearranged to obtain an equation for $\phi_C$ in terms of
$z_C$
\begin{equation}
\phi_C=z_C\exp\left(-\phi b\right).
\label{phic}
\end{equation}
Using Eq.~(\ref{flory_mix2}) to substitute for $f$ and
Eq.~(\ref{muc}) to substitute for $\mu_C$, in Eq.~(\ref{omeg}),
\begin{equation}
\omega=\frac{\phi}{n_B}\ln\phi+(1-\phi)\ln(1-\phi)-\phi_C,
\end{equation}
but we want it in terms of the relevant variables which are $\phi$
and $\mu_C$ so we use Eq.~(\ref{phic}) to substitute for $\phi_C$
\begin{equation}
\omega=\frac{\phi}{n_B}\ln\phi+(1-\phi)\ln(1-\phi)-
z_C\exp\left(-\phi b\right).
\label{omega}
\end{equation}
This equation completely describes the thermodynamics of
the mixture.

\subsection{The critical point}

We begin the determination of the phase behaviour by finding the
critical point for demixing.
First we need the chemical potential $\mu$ of the polymer, which is
just the $\phi$ derivative of Eq.~(\ref{omega}),
\begin{equation}
\mu= \frac{\ln\phi}{n_B}+n_B^{-1}-1
-\ln(1-\phi)+z_Cb\exp(-\phi b).
\end{equation}
Now, the critical point is the point where the first and second $\phi$
derivatives of the chemical potential are equal to zero. The
derivatives are
\begin{eqnarray}
\frac{\partial\mu}{\partial\phi}=\frac{1}{n_B\phi}
+\frac{1}{1-\phi}-z_Cb^2\exp(-\phi b)
\label{cpeq2a}\\
\frac{\partial^2\mu}{\partial\phi^2}=-\frac{1}{n_B\phi^2}
+\frac{1}{(1-\phi)^2}+z_Cb^3\exp(-\phi b).
\label{cpeq2}
\end{eqnarray}
Setting them to zero results in two simultaneous equations
for the polymer blob density
$\phi^{cp}$ and the sphere activity $z_C^{cp}$ at the critical point.
Combining these two equations yields
an equation solely in terms of $\phi^{cp}$,
\begin{eqnarray}
-(1-\phi^{cp})^2+n_B\left(\phi^{cp}\right)^2+
b\phi^{cp}(1-\phi^{cp})^2+&&\nonumber\\
bn_B\left(\phi^{cp}\right)^2\left(1-\phi^{cp}\right)
&=&0,
\label{phicp}
\end{eqnarray}
and also one for $z^{cp}_C$
\begin{equation}
z_C^{cp}=b^{-2}\left[\frac{1}{n_B\phi^{cp}}+\frac{1}{1-\phi^{cp}}\right]
\exp(\phi^{cp}b).
\label{zcp}
\end{equation}
Equation (\ref{phicp}) may be solved numerically for $\phi^{cp}$ and
then $z_C^{cp}$ obtained from Eq.~(\ref{zcp}).

For large $n_B$ the equations simplify and
we can solve the equations explicitly. For large $n_B$ we
look for a solution with $\phi^{cp}$ small. Eqs. (\ref{phicp})
and (\ref{zcp}) then yield
\begin{equation}
\phi^{cp}=\frac{n_B^{-1/2}}{\sqrt{(1+b)}}~~~~~~
z_C^{cp}=\frac{1}{b^2}\left(1+n_B^{-1/2}2{\sqrt{1+b}}\right)
~~~~~n_B\gg1.
\label{cpsaw}
\end{equation}
For large polymers, $n_B\gg1$, at the critical point
the density of polymer blobs scales as $n_B^{-1/2}$ while
the sphere activity tends to a constant as $n_B$ increases.
The reduced density of spheres
$\phi_C$ tends to $b^{-2}$ for $n_B$ large, from Eqs.~(\ref{phic})
and (\ref{cpsaw}).

In order
to estimate the magnitude of the sphere-sphere interactions we require the
volume fraction $\eta$ of the spheres,
$\eta=\rho_C(\pi/6)\sigma^3=\rho_C(B_{SB}/8)$.
At the critical point the volume fraction of the spheres is
\begin{equation}
\eta^{cp}=\frac{1}{8b}\left(1+n_B^{-1/2}\frac{2+b}{\sqrt{1+b}}\right)
~~~~~n_B\gg1.
\label{etacp}
\end{equation}
For large polymers the volume fraction $\eta$ of spheres at the
critical point is close to $1/8b$.
For our neglect of the sphere-sphere interactions to be
valid $\eta$ must be small. We estimate $b$ in the next section and
find it to be generally around
4 or larger, so the volume fraction of spheres at
the critical point is roughly $0.03$ or less and our neglect
of sphere-sphere interactions is not unreasonable.
Having defined the volume fraction of spheres we can also define
an effective volume fraction of the blobs. If we regard each blob
as filling a spherical volume of diameter $\sigma$ then the
`volume fraction' of blobs equals $\phi(b/8)$ and so at the
critical point we have a blob volume fraction at the critical point
\begin{equation}
\eta_B^{cp}=\frac{n_B^{-1/2}b}{8\sqrt{(1+b)}}
~~~~~n_B\gg1.
\label{etabcp}
\end{equation}

\begin{figure}[t]
\begin{center}
\caption{
\lineskip 2pt
\lineskiplimit 2pt
The variation of the volume fractions of polymer blobs and
colloidal spheres at the critical demixing point, as a function
of polymer length $n_B$.
The solid and long-dashed curves are
for polymer in a good solvent, $b=3.9$; the solid curve is the
volume fraction of polymer blobs, $\eta_B^{cp}$ and the long-dashed curve
curve is that of the spheres, $\eta^{cp}$.
The dotted and dot-dashed curves are
for polymer in a quite poor solvent, $b=20$; the dotted curve is
$\eta_B^{cp}$ and the dot-dashed curve is $\eta^{cp}$.
\label{pdnb}
}
\vspace*{0.1in}
\epsfig{file=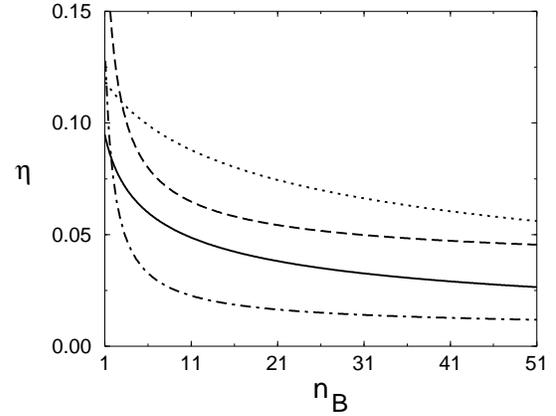,width=3.0in}
\end{center}
\end{figure}

\begin{figure}[t]
\begin{center}
\caption{
\lineskip 2pt
\lineskiplimit 2pt
The phase diagrams of two colloid-polymer mixtures in the $\eta$-$\eta_B$
plane; the $x$ and $y$ axes are the volume fractions of the
colloidal particles and blobs, respectively. The curves denote the
coexisting densities and the black circles denote the critical points.
Both curves are for $n_B=5$ blobs.
The solid curve is for a good solvent $b=3.9$ while the long-dashed
curve is for a rather poor solvent, $b=20$. The dotted lines are tie lines,
lines connecting two coexisting phases.
\label{pd}
}
\vspace*{0.1in}
\epsfig{file=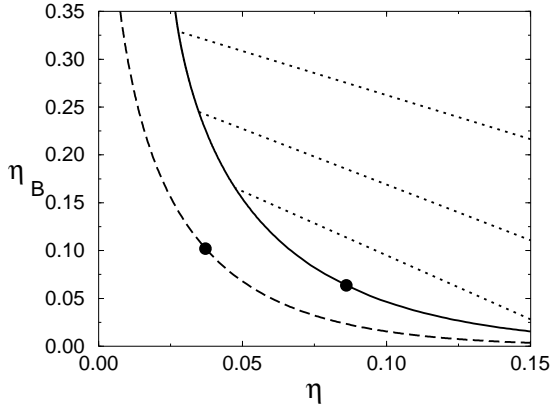,width=3.0in}
\end{center}
\end{figure}

In order to illustrate the trends in the demixing behaviour with
polymer length, measured by $n_B$, we have calculated
(using Eqs.~(\ref{phicp}) and (\ref{zcp}))
the volume fractions of spheres and of blobs at the
critical demixing point, $\eta^{cp}$ and $\eta_B^{cp}$, respectively.
The results are shown in Fig.~\ref{pdnb} where the solid curve is
the volume fraction of blobs, and the long-dashed curve is the sphere
volume fraction. For large $n_B$, we see that while the
sphere volume fraction is tending towards a plateau, that of the
polymer blobs is continuing to decrease, which is just what
we expect from Eqs.~(\ref{etacp}) and (\ref{etabcp}).
For the calculations we have set $b=3.9$ which is approximately
its value in the $a/\sigma\rightarrow0$, $B_M\ne 0$ limit.
We will discuss the estimation of $b$ and its variation with solvent
quality in section \ref{secsq}. The phase diagram in the
plane of the volume fractions of colloidal spheres and blobs,
for spheres and polymers with $n_B=5$ blobs is
plotted in Fig.~\ref{pd}. Note the large region of fluid-fluid
coexistence.

\subsection{Virial expansion}

We can expand out our semigrand potential Eq.~(\ref{omega}) as
a virial or density expansion
\begin{eqnarray}
\omega  &=&\frac{\phi}{n_B}\ln\phi-
z_C-\phi+zb\phi+\frac{1}{2}\left(1-z_Cb^2\right)\phi^2+\nonumber\\
&&\frac{1}{6}\left(1+z_Cb^3\right)\phi^3+
\frac{1}{12}\left(1-\frac{1}{2}z_Cb^4\right)\phi^4+\cdots .
\nonumber\\
\label{vir_saw}
\end{eqnarray}
So, the second $B_2$
and third $B_3$ virial coefficients for the monomer-monomer
interactions in the presence of the colloidal spheres are
\begin{equation}
\begin{array}{cc}
B_2=\frac{1}{2}\left(1-z_Cb^2\right)(2B_B) &
 B_3=\frac{1}{6}\left(1+z_Cb^3\right)(2B_B)^2.
\end{array}
\label{b2s}
\end{equation}
Using Eq.~(\ref{cpsaw}) for the activity
of spheres at the critical point in the above equation for the
virial coefficients, we find that the
critical point occurs when the second virial coefficient in the expansions
is negative and small, of order
$B_Bn_B^{-1/2}\sim\sigma^3n_B^{-1/2}$, when $n_B$ is large.
This holds for $B_B$ of order $\sigma^3$ which is correct when
the excluded volume interactions within a blob are strong, see
the next section for a discussion of this point.
The third virial coefficient is of order $\sigma^6$ at the critical
point, again assuming the intra-blob excluded volume interactions are
strong.
This is just as in the standard
Flory-Huggins free energy for a polymer in a poor solvent \cite{degennes}.

Thus, within our simple mean-field theory,
adding small colloidal particles, $\sigma\ll R_E$, to large
polymer molecules in a good solvent is essentially equivalent to
altering, worsening,
the solvent quality. As more and more spheres are added
the rescaled monomers, the blobs, start to attract each other and so
the polymer and spheres demix just as polymer and a poor solvent
demix. If this picture is correct then at length-scales large in
comparison to the blob size $\sigma$ the polymers will behave
just as a normal polymer in a good, theta or poor solvent, depending
on the concentration of spheres.
The behaviour of polymers as the solvent quality
is worsened and phase separation occurs has been well-studied and
is now well understood; see the results of recent computer simulations
\cite{frauenkron97,wilding96,yan00}. These simulations have found that the
mean-field theory prediction of 
$n_B^{-1/2}$ scaling of the density of polymer at the critical point
and the size of the second virial coefficient $B_2$, are
almost correct, there
are only logarithmic corrections.

\section{Solvent quality}
\label{secsq}

The phase behaviour
depends on only two parameters: the number of blobs of
size $\sigma$ in the polymer, $n_B$, and the ratio of the blob-sphere
to blob-blob excluded volume, $b$. The blob-blob excluded volume, $2B_B$,
is needed to convert from our reduced densities to number densities,
but only the ratio $b=B_{SB}/(2B_B)$ affects the nature of the phase behaviour.

First, let us consider polymers in which the excluded volume interactions
are strong, we are far from the theta temperature, and the particles are not
too small. This is the `excellent' solvent regime of Odijk
\cite{odijk00}. Note that whether or not a solvent is excellent
in this sense depends not only on the properties of the solvent but
on the size of the particles.
There is a parameter which describes how strong the
excluded volume interactions are \cite{degennes,schaefer}, it is often
denoted by $\zeta$. For a single blob we have $\zeta_B$ which is
\cite{degennes,schaefer}
\begin{equation}
\zeta_B= \frac{B_M}{a^3}N_B^{1/2}.
\label{zetab}
\end{equation}
When $\zeta_B\ll1$ then an individual blob is close to being a random
walk --- the excluded volume interactions within a single blob are
negligible. In the other limit, $\zeta_B\gg1$ the excluded volume
interactions within a single blob are strong.

\begin{figure}[t]
\begin{center}
\caption{
\lineskip 2pt
\lineskiplimit 2pt
The variation of the volume fractions of polymer blobs and
colloidal spheres at the critical demixing point, as a function
of solvent quality, measured by $b$.
The solid and long-dashed curves are
for a polymer of length $n_B=3$;
the solid curve is the
volume fraction of polymer blobs, $\eta_B^{cp}$ and the long-dashed curve
curve is that of the spheres, $\eta^{cp}$.
The dotted and dot-dashed curves are
for a polymer of length $n_B=30$; the dotted curve is
$\eta_B^{cp}$ and the dot-dashed curve is $\eta^{cp}$.
\label{pdb}
}
\vspace*{0.1in}
\epsfig{file=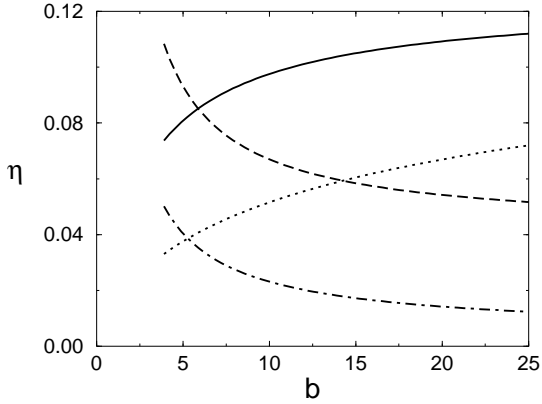,width=3.0in}
\end{center}
\end{figure}

The $\zeta_B\rightarrow\infty$ limit is the limit of a large SAW,
as $N_B\rightarrow\infty$, $\zeta_B\rightarrow\infty$, Eq.~(\ref{zetab}).
This is the scaling regime of an SAW, which is widely studied
and employed, it is the limit in which $R_E$ scales as a power law
of $N$ with the Flory exponent $\nu$ which is close to $3/5$.
Calculations on SAWs \cite{schaefer,nickel91} give the
second virial coefficient between two SAWs with radii
of gyration $R_G$ as $B_B=5.5R_G^3$ in the large $\zeta_B$ limit.
They also find $R_E=2.5R_G$, so $B_B=0.35\sigma^3$
between two SAWs with mean end-to-end
separations of $\sigma$.
Hanke {\it et al.} \cite{hanke99} have
applied field theory to obtain the result that $B_{SB}=2.7\sigma^3$
\cite{calc}. This result is not exact but is more than accurate
enough for the purposes of the present theory. This result
is obtained in the scaling limit of strong excluded volume interactions.
For comparison, for ideal polymers $B_{SB}=3.0\sigma^3$.
A swollen blob is more open than one which is ideal so $B_{SB}$
is correspondingly smaller for polymers with the same $R_E$.
So, with strong excluded volume interactions
even within blobs the ratio $b=3.9$. In this limit the blob-blob
interaction is strong and the chain as a whole will be swollen so
if the radius $R_E$ of the chain is known then $n_B$ may be estimated
from $n_B\sim(R_E/\sigma)^{5/3}$. Alternatively, if the number
of monomers $N_B$ in a chain with $R_E=\sigma$ is known then the
number of blobs may be found from $n_B=N/N_B$.

So far we have considered only polymer-colloid mixtures in which the
solvent for the polymer is sufficiently good and the colloidal
particle sufficiently large that $\zeta_B\gg1$ and even pieces
of the polymer as small as $\sigma$ are strongly swollen. Then we
can use the value of the blob-blob and blob-sphere second virial coefficients,
$B_B$ and $B_{SB}$,
in the $\zeta_B\rightarrow\infty$ limit. But what if the solvent
quality is less good and the particles not too large?
As the solvent quality decreases
the monomer-monomer interaction, $B_M$ decreases from its value
in a good solvent which is of order $a^3$. This will decrease the
blob-blob interaction, measured by $B_B$, while leaving the
blob-sphere second virial coefficient $B_{SB}$ still at around $\sigma^3$.
The second virial coefficient for the interaction between an ideal
chain, $R_E=\sigma$, and a hard sphere, of diameter $\sigma$,
is known exactly \cite{eisenriegler96}
and is close to $3\sigma^3$. Thus, when $\zeta_B$ is no longer much
larger than one, the ratio between the sphere-blob and blob-blob
excluded volumes, $b$, will increase as $\zeta_B$ decreases.
It is divergent for ideal
polymers as then the blob-blob excluded volume is zero.

We expect that as the solvent quality for the polymer worsens and the
polymer-polymer interactions weaken that phase separation will be
enhanced, the polymer and colloid will be less miscible. With this
in mind we return to the equation for the density of polymer at
the critical point, Eq.~(\ref{phicp}). We assume that the critical
density will be very low but make no further assumptions,
Eq.~(\ref{phicp}) then simplifies to
\begin{equation}
-1+n_B\left(\phi^{cp}\right)^2+
b\phi^{cp}+bn_B\left(\phi^{cp}\right)^2 =0,
\label{phicp2}
\end{equation}
which is a quadratic solution with a physical root
\begin{equation}
\phi^{cp}=\frac{-1+\sqrt{1+4(n_B/b)(1+1/b)}}{2n_B(1+1/b)}.
\label{phicp3}
\end{equation}
For $n_B/b$ large, i.e., long polymers with blob-blob interactions
which are not too weak, this equation simplifies to Eq.~(\ref{cpsaw})
--- for very long polymers even weak monomer-monomer and hence
blob-blob excluded volume interactions are sufficient to obtain the
standard scaling of the critical density as $n_B^{-1/2}$.
The other limit in Eq.~(\ref{phicp3}) is when $n_B/b$ is small,
i.e., the blob-blob interactions are weak and the polymer is not
too long. In this limit, Eq.~(\ref{phicp3}) simplifies to
\begin{equation}
\phi^{cp}=b^{-1}~~~~~~n_B/b\ll1,
\label{cprw1}
\end{equation}
or
\begin{equation}
\eta_B^{cp}=1/8~~~~~~n_B/b\ll1.
\label{cprw2}
\end{equation}
The density of blobs at the critical point is independent of the
length of the polymer. Recalling that $\phi$ is the number density
of blobs times $2B_B$, we rewrite Eq.~(\ref{cprw1}) as
\begin{equation}
\rho_P^{cp}=\frac{1}{n_BB_{SB}}~~~~~~n_B/b\ll1,
\label{cprw}
\end{equation}
where $\rho_P^{cp}$ is the number density of polymer molecules
at the critical point, which is of order $1/(n_B\sigma^3)$.
This scaling has been derived before for mixtures of colloidal
hard spheres and ideal polymers \cite{sear01}.
Our present theory for polymers with excluded volume interactions,
although rather different in a number of ways from the theory
of Ref.~\cite{sear01} predicts the same critical density of polymer
when these excluded volume interactions are turned off. Giving
us confidence in both theories. We can easily obtain the
density of spheres at the critical point in the $n_B/b\ll1$ limit.
The reduced activity is $z_C^{cp}=e/(n_Bb)$, from Eq.~(\ref{zcp}), and
hence the reduced density at the critical point $\phi_C^{cp}=1/(n_Bb)$ and
finally the volume fraction
\begin{equation}
\eta^{cp}=\frac{1}{8n_B}~~~~~n_B/b\ll1,
\label{etarw}
\end{equation}
which again is the same scaling as found previously \cite{sear01}.
For weak interactions between the blobs the density of spheres at
the demixing critical point is dramatically below its value when there
are strong interactions between the blobs, Eq.~(\ref{etacp}).
Also, note that the $\zeta$ parameter for the whole chain is
$\zeta_P=N^{1/2}B_M/a^3\simeq n_B^{1/2}B_B/\sigma^3$. We
can rewrite this in terms of $b$, $\zeta_P\simeq n_B^{1/2}/b$,
ignoring a numerical prefactor. When $n_B/b\ll1$, then substituting
for $b$, $\zeta_P\ll n_B^{-1/2}$ and so as $n_B$ is larger
than one $\zeta_P\ll1$ --- an individual polymer molecule in dilute
solution and before the colloidal spheres are added is ideal,
its $R_E=aN^{1/2}$. Equations (\ref{cprw1}) to (\ref{etarw}) are for
demixing of a polymer which is effectively ideal, its monomer-monomer
excluded-volume interactions are negligible.

Returning to Fig.~\ref{pdnb}, we can compare the densities at
the demixing critical
point for polymers in a good solvent (solid and long-dashed
curves) with those in a rather poor solvent, $b=20$ (dotted and
dot-dashed curves). We see that as the solvent quality worsens the
volume fraction of polymer blobs increases (compare the solid and
dotted curves) and the volume fraction of spheres decreases (compare
the long-dashed and dot-dashed curves). Also, for $b=20$ and for
not-too-large $n_B$ the density of blobs at the critical point
for demixing is relatively insensitive to $n_B$, which is
what we expect from Eq.~(\ref{cprw2}). Note that for the polymer in
a good solvent the effective volume fraction of polymer blobs
is lower than that of the spheres at the critical point whereas
for a poor solvent, $b=20$, the opposite is true.

In Fig.~\ref{pdb} we have plotted the
the variation of the densities at the critical
point for fixed $n_B$ and varying solvent quality $b$.
For a small polymer, $n_B=3$, the volume fractions
of blobs and spheres are plotted as the solid and long-dashed curves,
respectively,
while the dotted and dot-dashed curves are the volume fractions
of blobs and spheres, respectively, for a much longer polymer,
$n_B=30$. The mixture of the longer polymer and the spheres demixes
at lower densities than the mixture of the shorter polymer and
spheres of course, and as the solvent quality decreases the 
volume fraction of polymer increases while that of the spheres
decreases. In Fig.~\ref{pd} we have plotted the phase diagrams in the
plane of the two volume fractions for polymers of length $n_B=5$
for a good solvent, $b=3.9$
and a rather poor solvent, $b=20$.
The coexistence curve for the poorer solvent lies outside that for the
good solvent: reducing the solvent quality increases the extent of the
immiscibility.

The larger density of polymer at the critical point comes from the
fact that, near their critical points,
the third
virial coefficient is relatively much larger for ideal polymers than
for polymers with excluded volume interactions.
The values of the virial coefficients are obtained by inserting the
activity of colloidal spheres at the critical point, $z_C^{cp}$,
into Eq.~(\ref{b2s}).
For polymers with excluded volume interactions near the critical
point $B_3=O(\sigma^6)$. While for ideal
polymers near their critical point $B_3=O(\sigma^6/n_B)$. Small
$B_3$'s lead to high critical densities, see the appendix for details.

\subsection{Comparison with computer simulation}

Finally in this section,
we compare with the results of recent computer simulations
of colloidal particles and hard spherical particles
by Meijer {\it et al.} \cite{three}.
They studied colloidal particles + SAWs with $R_E/\sigma=4.8$, $7.0$
and $9.9$. In the simulations the size of colloidal particle not that
of the polymer was varied but for the moment we will assume that in each
case $\zeta_B\gg1$ so that in the simulations
the blob-blob and blob-sphere interactions are both in the good solvent
scaling regime and our parameter $b=3.9$. This leaves us with
the problem of estimating the values of our parameter $n_B$ for
the simulated systems. For large polymers we must have that
$n_B\sim (R_E/\sigma)^{1/\nu}$ as $n_B$ is extensive in the contour
length of the polymer. In order to obtain an estimate for $n_B$ we set the
unknown numerical prefactor in this scaling relation to 1, and
$\nu=0.6$, and so obtain
$n_B=14$, 26 and 46 for the three simulated systems.

For these three systems we predict critical points at
polymer blob volume fractions $\eta_B^{cp}=0.045$, $0.035$
and $0.028$, respectively. Preliminary simulation results for
the critical densities are
$\rho_P^{cp}/\rho_P^* =2.04$, $3.19$ and $4.65$, where
$\rho_P^*=1/[(4\pi/3)R_G^3]$.
Using the theoretical result $R_E=2.5R_G$ in the good solvent regime and
converting from $R_E/\sigma$ to $n_B$ as above we have that
$\rho_P=3.7/n_B^{1.8}$. Converting our theoretical predictions to
values of the ratio $\rho_P^{cp}/\rho_P^*$, we have
$\rho_P^{cp}/\rho_P^* =0.19$, $0.25$ and $0.31$. The theoretical
predictions are about an order of magnitude too small, although the trend
with increasing $R_E/\sigma$ is correct. Part of the discrepancy may
come from our crude estimation of the relationship between the size
of the polymer, $R_E/\sigma$, and the number of blobs, $n_B$, but it seems
very likely that the theory is also underpredicting the density of polymer
when it demixes from the spheres. Preliminary results for the volume
fraction of spheres at the critical demixing point, $\eta^{cp}$,
are around $0.2$-$0.25$ whereas we find $0.03$-$0.05$ for this range
of values of $R_E/\sigma$. Again the theory overestimates the extent
of the immiscibility. It should be noted that, in simulation,
for the largest value of $R_E/\sigma$, the
colloidal spheres have a diameter only about ten times that of the monomer.
If the blobs are too small to be in the good solvent regime, the requirement
$\zeta_B\gg 1$ will be violated, and our parameter
$b$ will be $>3.9$. Then
the weaker blob-blob interactions in simulation will increase
the polymer density at demixing with respect to that given by the
theoretical prediction for the good-solvent scaling regime.
The simple theory derived here is clearly not quantitative but this
is perhaps no surprise, it is really only capable of giving rough estimates
and the qualitative nature of trends. The precise nature of the trends
for $R_E/\sigma\gg1$ will be those for a long polymer in a poor solvent;
see the simulation results of Refs.~\cite{frauenkron97,wilding96,yan00}.

\section{Conclusion}

Mixtures of hard spheres and larger, flexible polymers which do not absorb
onto the surface of the spheres exhibit extensive immiscibility. The
cross excluded volume interactions in these mixtures,
i.e., the excluded volume interaction between the sphere and the
polymer, are large. So, the spheres and polymers `get in each others way'
so reducing each others entropy and driving them apart into separate
phases \cite{frenkel92}.
The tendency to demix increases as the polymers become
larger and larger, because the excluded volume interactions scale
linearly with the length of the polymer while the translational entropy
gained by mixing solutions of spheres and polymers does not vary with
polymer length. Here we derived a simple analytic theory for these mixtures
and found that when the excluded volume interactions were so strong that
even blobs of size equal to that of the spheres were swollen,
a mixture of polymer and much smaller spheres behaves much as a polymer
in a poor solvent does. By contrast, as we showed in earlier work, a mixture
of spheres and ideal polymers behaves rather differently. There the
effective third virial coefficient of the polymer is very small, which
pushes up the polymer density at the critical point for demixing.

Our results are of relevance to mixtures of globular
proteins and polymers, as in these mixtures it is easily possible
to have polymers large than the protein. However, our assumption of a
purely repulsive interaction between the polymer and the spheres is
rather unrealistic for proteins which
have rather complex surfaces. Some part of this complex surface
may well attract the monomers.
A clean comparison with experiment could
be done with experiments on either small synthetic colloidal particles
or nanoparticles. Another possibility is to instead of making the
colloidal particles smaller, make the polymer bigger
by employing DNA \cite{verma00}.
There however, the colloidal particles would have to be reasonably
large as the effective monomer length of DNA is about $a=100$nm and our
theory assumes that $a\ll\sigma$. An objective for future work could
be to relax this restriction to account more accurately
for mixtures of spheres with
semiflexible polymers such as DNA.

It is a pleasure to thank P. Bolhuis, A. Louis and E. J. Meijer
for inspiring conversations, and for sending me results
of their computer simulations, and
A. Hanke for providing me with
his calculated value for $B_{SB}$. I would also like to thank D. Frenkel,
also for inspiring conversations, and
for the invitation to visit AMOLF, where this work was started,

\section*{Appendix: Virial expansions and critical densities}

Here we explore how the critical density varies with the size of the
third virial coefficient. Consider the simplest possible virial expansion
that has a critical point, an expansion truncated after the third
virial coefficient. For the chemical potential $\mu$ as a function of density
$\rho$ this is
\[
\mu=\ln\rho+2B_2\rho+3B_3\rho^2,
\]
where $B_2$ and $B_3$ are the second and third virial coefficients,
respectively.
Assuming that $B_3$ is fixed and that $B_2$ varies linearly with some
temperature-like variable $t$, we have $B_2=B(1-t)$, and
\[
\mu=\ln\rho+2B(1-t)\rho+3B_3\rho^2.
\]
The critical point occurs when the first and second derivatives
of the chemical potential are zero, giving two equations for the critical
value of $t$, $t^{cp}$, and the critical density $\rho^{cp}$
\begin{eqnarray*}
\frac{1}{\rho^{cp}}+2B(1-t^{cp})+6B_3\rho^{cp} &=&0\\
-\frac{1}{\left(\rho^{cp}\right)^2}+6B_3 &=&0
\end{eqnarray*}
The second equation gives the density at the critical point straightaway,
$\rho^{cp}=1/\sqrt{6B_3}$: the critical density does not depend on the
value of the second virial coefficient but on that of the third virial
coefficient. For our mixtures of ideal polymers and spheres, as the
polymers are ideal their blobs do not repel each other, they only
interact with each other via the spheres. Thus the third virial
coefficient of the polymer blobs (at constant sphere chemical potential)
is very small and the density at the critical point of demixing
correspondingly high.


\begin{thebibliography}{99}
\bibitem{notesg} Note that our semigrand potential is a function
of the density of polymer and the chemical potential of spheres,
whereas the semigrand potential of mixtures of spheres with
small polymer molecules of Ref.~\cite{lekkerkerker92} and the
body of work which has followed this reference, is a function
of the chemical potential of the polymer and the density of spheres.
In principle of course it is possible to develop exact theories
for both semigrand potentials. However, in practice the component
whose density is specified is the one whose interactions cannot be
neglected. When the spheres are larger than the polymer molecules,
the sphere-sphere interactions are important as demixing tends
to occur at high sphere volume fractions. When the polymer molecules
are larger the blob-blob interactions are important as we show
in section \ref{secsq}.
\end{thebibliography}
\end{document}